\documentclass{article}
\usepackage{graphicx} 
\usepackage{amsthm,amsmath, amssymb}
\usepackage{todonotes}
\usepackage[numbers]{natbib}

\newcommand{\mbL}{\mathbb{L}}

\newcommand{\mbR}{\mathbb{R}}

\newcommand{\mcF}{\mathcal{F}}

\newcommand{\mcM}{\mathcal{M}}

\newcommand{\diff}{\mathrm{d}}

\usepackage{longtable,booktabs,array,calc}
\usepackage{hyperref}

\theoremstyle{plain}\newtheorem{theorem}{Theorem}[section]
\theoremstyle{plain}\newtheorem{corollary}{Corollary}[section]
\theoremstyle{definition}\newtheorem{definition}{Definition}[section]

\title{Comparison of SRVF and Signature Kernels Analyzing Chromosomal Conformations}
\author{Jose M. Cruz Mendez \\ 
Independent Researcher\\
\and 
Carlos J. Soto\\
Department of Mathematics and Statistics\\
University of Massachusetts Amherst \\
}
\date{}

\begin{document}

\maketitle

\begin{abstract}
Comparing populations of three-dimensional curves by their shape is a common task in statistical shape analysis, as structure and function are intimately related. Elastic shape analysis based on the square-root velocity function (SRVF) is a state of the art approach for this task as it has many desirable properties. Signature kernels compare curves through their iterated integrals and are invariant to translation and reparametrization by construction, but they have not been used to compare populations of shapes. We extend their invariance to scale and rotation, consider several readouts of the PDE that computes them, and use the resulting distances in downstream analysis. As a baseline, we compare our results to those achieved by utilizing the SRVF framework. We demonstrate the robustness of the signature readouts on synthetic helices across several noise levels, and show the lack of robustness for elastic shape analysis. 
We then compare chromosome conformations reconstructed from high-throughput chromosome conformation capture (Hi-C) data of wild-type cells and of two knock-out conditions, both over whole regions and within their topologically associating domains (TADs).
\end{abstract}

\section{Introduction}
Classification and quantification of similarity of three-dimensional curves is a pervasive endeavor in machine learning and statistical analyses. 
The classical $\mbL^2$ distance with subsequent statistical analysis is the most standard approach, however inherently this dissimilarity measure does not capture \textit{shape} differences. Shape analysis is an area which defines similarity of shape according to some \textit{invariances}, e.g. rotation, scale, location, and diffeomorphisms. That is, these operations change the representation of the shape, but not the shape itself.
It is generally straightforward to remove dependence on these nuisance parameters other than diffeomorphisms, or time warping. That is, the time points at which curves are observed play a role in similarity measures. Many methods have been proposed to remove this effect, for instance pre-processing via arc-length parameterization or enforcing a penalty term on the diffeomorphism. Another method to remove this dependence is to build a metric which itself incorporates the invariances. The most prolific to take this approach is the \textit{elastic shape analysis}\citep{srivastava2016functional} framework which builds upon the Fisher-Rao metric\citep{rao1945information}.

A recent advancement in constructing invariant distances is signatures. Signatures have been used to classify curves up to reparametrization \citep{celledoni2019signatures} and extended to paths in Lie groups such as $SO(3)^n$ \citep{lee2020pathsignatures}. The signature kernel yields a two-sample test for laws of stochastic processes \citep{chevyrev2022signature}. To our knowledge, signature kernels have not been used to compare populations of shapes, nor applied to chromosome conformations. 
In the following we demonstrate the robustness and applicability of signature kernels on synthetic and real data. As a baseline we compare our method to the elastic shape analysis framework. Both of these methods are invariant to reparameterization, however, we show that our method is more robust to noise, as \citet{Feng2010-bz} found for integral invariants of curves. Lastly, for real data we consider three-dimensional conformations and consider them at localized, segmented levels and at the larger scale. 

\section{Background}
In the following we discuss some of the background on the state of the art techniques for analyzing three-dimensional curves. We first introduce elastic shape analysis and signature kernels and refer to \cite{srivastava2016functional} and \cite{chevyrevPrimerSignatureMethod2025} for more complete expositions.

\subsection{Elastic Shape Analysis}\label{sec:elastic-shape-analysis}
Let $\mcF$ be the set of absolutely continuous functions $\mcF\ni f:[0,1]\rightarrow \mbR^3$. Elastic shape analysis takes into account the geometry of the space of curves irrespective of parameters which do not effect shape, i.e., size, location, rotation, and parameterization. These artifacts of shape have a correspondence with algebraic groups and their action on the set of functions is something we desire invariance to. Formally, let $(\Gamma,\circ)$ be the group of of diffeomorphisms $\Gamma\ni g:[0,1]\rightarrow[0,1]$  equipped with composition, $(SO(n),\times)$ be the group of $n\times n$ special orthogonal (rotation) matrices, and $(\mbR^n,+)$ be the group of vector translations, and $(\mbR^+,\cdot)$ be the group of scalars.

We wish to construct a distance or metric which is invariant to the actions of the above groups on the space of functions. The $\mbL^2$ distance is a natural candidate and the starting point. Invariance to size and location are simple pre-processing steps. First, remove location by setting $f\mapsto f-\int \diff t f $ then remove dependence on size by setting $f\mapsto f/\|f\|_2$. Moving forward, for elastic analysis we assume the functions are centered at the origin and scaled to be unitary. The $\mbL^2$ distance of these centered and scaled functions is thus $d(f_1,f_2)=\|f_1-f_2\|_2$ and is invariant to size and location by construction. To construct a distance invariant to rotational differences we have rotational alignment as $$\min_{O_1,O_2\in SO(n)}d(O_1f_1,O_2f_2)=\min_{O\in SO(n)}d(f_1,Of_2).$$ One could then attempt to make the $\mbL^2$ distance invariant to diffeomorphisms in a similar manner i.e. $\min_{\gamma\in\Gamma}d(f_1\circ\gamma,f_2)$ however this is not invariant to this action. That is, simple calculations show that $$\|f_1-f_2\|\ne\|f_1-f_2\circ\gamma\|$$ and further $$\|f\|\ne\|f\circ\gamma\|.$$ 
A common approach to circumvent this issue is to use the arc length parameterization, which is often satisfactory but has shortcomings.

Another manner to gain invariance is to utilize a metric other than the one induced by the $\mbL^2$ inner product, namely the Fisher-Rao metric\cite{rao1945information}. The Fisher-Rao metric is defined on the manifold of parameterized curves so at the tangent space of $f$, $T_f\mcM$, the inner product is 
$$\langle v,w\rangle_f=\frac{1}{4}\int \dot{v}(t)\dot{w}(t)\frac{1}{|\dot{f}(t)|}\diff t.$$ 
One can verify that this metric is invariant to reparametrization; location and scale are removed. Thus the distance we wish to utilize on the space of functions is the Fisher-Rao metric with location and scale removed as a pre-processing procedure. Unfortunately, though,  computing this distances is very complex due to the of the space and lack of a closed form geodesic equation.

Elastic shape analysis introduced the square-root velocity function (SRVF) transformation to circumvent the complex calculations ... from using the Fisher-Rao metric. The SRVF is the transformed function defined as $$Q(f)(t):=\frac{\dot{f}(t)}{\sqrt{\|\dot{f}(t)\|}}.$$ This transformation embeds functions onto an infinite-dimensional Hilbert sphere as $\|Q(f)\|_2=1$ and more remarkably under this transformation the Fisher-Rao metric reduces to the $\mbL^2$ distance. Thus, the distance is $$d(f_1,f_2)=\arccos\langle Q(f_1),Q(f_2)\rangle_2.$$
To gain invariance of rotation and diffeomorphisms one can now search over these spaces in an invariant manner. 
 $$d(f_1,f_2)=\min_{\gamma\in\Gamma,O\in SO(3)}\arccos\langle Q(f_1),O(Q(f_2),\gamma)\rangle_2$$
where $(Q(f),\gamma)$ is the corresponding group action in the SRVF space. Optimal rotation is dependent on the parameterization of the curve, so the above distance needs to be found iterating between optimization over rotations and optimization over diffeomorphisms.

\subsection{Signature Kernels}\label{sec:signature-kernels}

We first introduce signatures, from which signature kernels are built.

\subsubsection{The Signature Transform}\label{sec:signature-transform}

Let $X(t)$ be a path in $d$ dimensions with $t\in[0,1]$. Assume $X$ has finite total variation, that is, the supremum over all partitions $0 \leq t_1 \leq \cdots \leq t_J \leq 1$ of $\sum_{i=1}^{J-1} \|X(t_{i+1}) - X(t_i)\|$ is finite. Although finite total variation paths aren't necessarily differentiable, we limit ourselves to (piecewise) differentiable paths.

\begin{definition}[{\cite{chevyrevPrimerSignatureMethod2025}}]\label{def:path-signatures}
Let $X: [a,b]\mapsto \mathbb{R}^{d}$ be a piecewise differentiable path, which can be split into coordinates $(X_{1}, X_{2} \cdots X_{d})$. Let $U$ be an ordered sequence of coordinate indices of $X$. The path signatures of $X$ are defined over all sequences of indices $U$, recursively, as \[
S^{U + i}_{[a,b]}(X) = \int_{a}^{b} S^{U}_{[a,s]}(X) X'_{i}(s)\diff s
\] where we define the initial signature $S^{\emptyset}_{[a,b]}(X)=1$ for all intervals $[a,b]$.

We refer to $S^{U}_{[a,b]}(X)$ as a \emph{level-$m$ signature} if $|U|=m$.

Unpacking this definition, level-1 signatures become \[
S^{i}_{[a,b]}(X) = \int_{a}^{b}S^{\emptyset}_{[a,s]}X'_{i}(s)\diff s = \int_{a}^{b}X'_{i}(s)\diff s = X_{i}(b)-X_{i}(a)
\] and level-2 signatures become \[
S^{(i,j)}_{[a,b]}(X) = \int_{a}^{b}S^{i}_{[a,s]}(X)X'_{j}(s)\diff s = \int_{a}^{b} \int_{a}^{s}X'_{i}(r)X'_{j}(s)\diff r \diff s
\] More generally, a level-$m$ signature with $U = (u_{1}, u_{2}, \cdots, u_{m})$ can be written as: \[
S^{U}_{[a,b]}(X) = \int_{a < t_{1} < t_{2} < \cdots < t_{m} < b} \prod_{j=1}^{m} X'_{u_{j}}(t_{j}) \diff t_{1}\cdots \diff t_{m}
\]
\end{definition}

\begin{definition}[{\cite[Def. 1.4]{chevyrevPrimerSignatureMethod2025}}]\label{def:signature-transform}
The signature transform, denoted by $S_{[a,b]}(X)$, maps a $d$-dimensional path to the ordered infinite sequence of all Path Signatures of $X$ as \[
X \mapsto (1, S^{1}_{[a,b]}(X), S^{2}_{[a,b]}(X), \cdots, S^{d}_{[a,b]}(X), S^{(1,1)}_{[a,b]}(X), S^{(1,2)}_{[a,b]}(X), \cdots)
\]
\end{definition}

The signature transform has some important properties:

\paragraph{It is translation invariant.} As signatures are defined via derivatives, $X$ and $X+v$ for a constant vector $v$ are mapped to the same signature.

\begin{proof}
As all signatures are integrals of products of derivatives, and derivatives kill constant functions, we have that $S_{[a,b]}(X+v) = S_{[a,b]}(X)$, since $\forall i: (X_{i}+v_{i})'(t) = (X_{i}' + v_{i}')(t) = X_{i}'(t) + 0 = X'_{i}(t)$.
\end{proof}

\paragraph{It is reparametrization invariant.}

\begin{theorem}[{\cite{chevyrevPrimerSignatureMethod2025}}]\label{thm:reparametrization-invariance}
Let $\psi$ be a surjective, continuous, non-decreasing function mapping $[\alpha,\beta]\to [a,b]$, and let $X: [a,b]\to \mathbb{R}^{d}$. Now define $\tilde{X} = X(\psi(t))$ for $t \in [\alpha,\beta]$. Then $S_{[\alpha,\beta]}(\tilde{X}) = S_{[\psi(\alpha),\psi(\beta)]}(X)$.
\end{theorem}

\begin{proof}
For differentiable $X$, we find that $\tilde{X}'_{i} = (X(\psi(t))'_{i} = \psi'(t)X'_{i}(\psi(t))$, and by $u$-substitution with $u = \psi(t), \diff u = \psi'(t)\diff t$ we recover the same integral for every individual signature, so $S^{U}_{[\alpha,\beta]}(\tilde{X}) = S^{U}_{[\psi(\alpha),\psi(\beta)]}(X)$.
\end{proof}

As signatures are reparametrization invariant, we can safely refer to all paths as continuous maps $X: [0,1]\to \mathbb{R}^{d}$, as any compact time domain can be reparametrized to $[0,1]$.

Signatures cannot distinguish a path that retraces its own steps: a path followed by its reversal has the same signature as the constant path. Up to translation and such retracings (\emph{tree-like equivalence}), signatures are universal \citep[Prop. 1.35]{chevyrevPrimerSignatureMethod2025}:

\begin{theorem}[{\cite[Prop. 1.35]{chevyrevPrimerSignatureMethod2025}}]\label{thm:universality}
Let $\mathcal{Y}$ be the space of paths in $\mathbb{R}^{d}$ up to a quotient by translations and retracing the same path (alternatively, take $\mathcal{Y}$ to be the space of paths augmented with a time channel and a base point). For any compact set $C \subseteq \mathcal{Y}$, any continuous function $g:C\to \mathbb{R}$ and any $\varepsilon > 0$, there exists a linear functional $F$ such that $\lvert F(S_{[a,b]}(X))-g(X) \rvert< \varepsilon$ for all $X \in  C$. In other words, there exists an integer $M\geq 1$, real numbers $\lambda_{1}, \lambda_{2}, \cdots, \lambda_{M}$ and words $U_{1}, U_{2}, \cdots, U_{M}$ such that for all $X \in  C$ we have that \[
\left|g(X) - \sum_{i=1}^{M}\lambda_{i} S_{[a,b]}^{U_{i}}(X) \right| < \varepsilon
\]
\end{theorem}

Universality can also be obtained without the quotient by adding time as an extra coordinate, but this makes the signature depend on the speed at which the path is traversed. We care about shape, not speed, so we do not add a time channel. The quotient costs us nothing in practice: chromosomes occupy physical space, so two different points along the chromosome cannot occupy the same position, and a curve that never revisits a point contains no retracing.

\subsubsection{The Signature Kernel}\label{sec:signature-kernel}

The \emph{signature kernel} $K(X,Y)$ is the inner product between signature transforms of two paths $X,Y:[0,1]\to \mathbb{R}^{d}$: \[
K(X,Y)=\left\langle S_{[0,1]}(X),S_{[0,1]}(Y)\right\rangle = \sum_{U \in \text{index sequences}} S^U_{[0,1]}(X) S^U_{[0,1]}(Y)
\] We can also parametrize the kernel on subpaths up to a stopping time as \[
K_{(s,t)}(X,Y) = \langle S_{[0,s]}(X), S_{[0,t]}(Y) \rangle
\] so that $K(X,Y)=K_{(1,1)}(X,Y)$. To calculate the signature kernel, it is possible to compute the first few levels of signatures and then perform a dot product in the truncated space. However, we can exploit a fact about the signature kernel to use more memory-efficient and faster-converging methods.

\begin{theorem}[{\cite[Thm. 2.5]{salviSignatureKernelSolution2021}}]\label{thm:goursat}
Let $X,Y:[0,1]\to \mathbb{R}^{d}$ be two differentiable paths. The signature kernel $K_{(s,t)}(X,Y)$ satisfies the following PDE: \[
\frac{ \partial^{2}K_{(s,t)} }{ \partial s \partial t } = \left\langle  \dot{X}(s), \dot{Y}(t)  \right\rangle K_{(s,t)} \qquad K_{(s,0)}=K_{(0,t)}=1
\] where $\langle \cdot ,\cdot  \rangle$ is the Euclidean inner product on $\mathbb{R}^{d}$, and $\dot{X}(s), \dot{Y}(t)$ the time derivatives of $X$ and $Y$ evaluated at $s$ and $t$ respectively.
\end{theorem}

We use the framework in \citep{tamayo-riosScalableSignatureKernel2025} to calculate signature kernels.

\section{Invariant Signature Kernels}\label{sec:method}

\subsection{Invariances and normalization}\label{sec:invariances}

We want to compare chromosomes by shape alone, so the comparison should not change when a path is translated, reparametrized, rescaled or rotated. We go through these in order, with rotation is left to the readouts below.

\paragraph{Translation.} Signatures are translation invariant, so any kernel built from them is too.

\paragraph{Reparametrization.} Since $S_{[0,1]}$ is reparametrization invariant, so is the signature kernel $K(X,Y)$. This is no longer true once we read the kernel at intermediate times, as the running signature is only equivariant:

\begin{corollary}[Equivariance]\label{cor:equivariance}
Viewing signatures as a function of the endpoint, $S(X;t) := S_{[a,t]}(X)$, they are reparametrization-\emph{equivariant}: for $t\in[\alpha,\beta]$, \[
S_{[\alpha,t]}(X\circ\psi) = S_{[a,\psi(t)]}(X), \quad\text{i.e.}\quad S(\tilde{X};t) = S(X;\psi(t)).
\] This is Theorem~\ref{thm:reparametrization-invariance} applied to $\psi$ restricted to $[\alpha,t]$, which maps onto $[a,\psi(t)]$.
\end{corollary}

\begin{figure}[t]
\centering
\includegraphics[width=\textwidth]{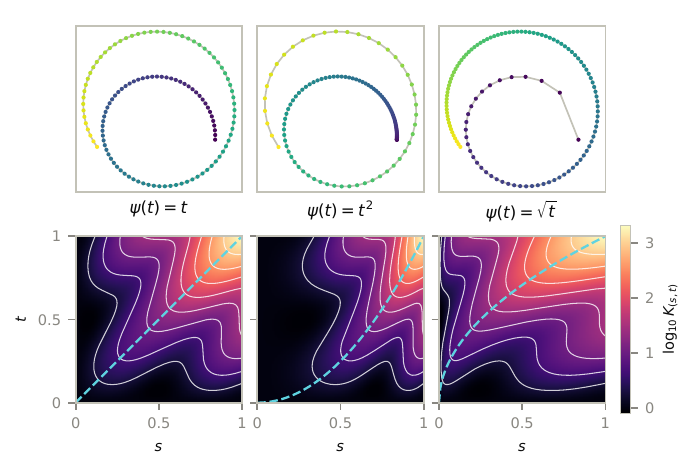}
\caption{Reparametrization equivariance of the PDE solution. Top: one curve $X$ sampled at $120$ equally spaced times under three parametrizations $X\circ\psi$, colored by time from dark to light. Bottom: $\log_{10}K_{(s,t)}(X\circ\psi,X)$ with isoclines, where $s$ is the time of $X\circ\psi$ and $t$ the time of $X$. By Corollary~\ref{cor:equivariance} this equals $K_{(\psi(s),t)}(X,X)$, so each panel is the first one with its horizontal axis re-timed by $\psi$. The dashed curve is $t=\psi(s)$. Both paths are normalized to total variation $L=8$.}
\label{fig:equivariance}
\end{figure}

Applying this to both paths, $K_{(s,t)}(X\circ\psi, Y\circ\phi) = K_{(\psi(s),\phi(t))}(X,Y)$. Figure~\ref{fig:equivariance} shows this for one curve under three parametrizations. We return to this when we discuss readouts.

\paragraph{Signatures and signature kernels are bounded.} Let $L_X = \int_0^1\|\dot X(t)\|_1 \diff t$ be the length of $X$, measured with the $\ell^1$ norm. The level-$m$ signatures satisfy $\sum_{|U|=m}|S^U(X)| \le L_X^m/m!$ \citep[Prop. 2.2]{lyonsDifferentialEquationsDriven2007}, so summing over all levels gives $\sum_U|S^U(X)| \le e^{L_X}$. A product of two sums of non-negative terms contains all of its diagonal terms, so \[
|K(X,Y)| \le \sum_U|S^U(X)| |S^U(Y)| \le \Big(\sum_U|S^U(X)|\Big)\Big(\sum_V|S^V(Y)|\Big) \le e^{L_X+L_Y}.
\] The same argument restricted to a single level gives $\left|\sum_{|U|=m}S^U(X) S^U(Y)\right| \le (L_XL_Y)^m/(m!)^2$. Everywhere else in this paper, length means total variation with the Euclidean norm. Since $\|v\|_1 \le \sqrt{d} \|v\|_2$, both bounds still hold with $L$ replaced by $\sqrt{d}$ times the Euclidean length, so the kernel grows at most exponentially in the length of the paths.

\paragraph{Scale.} Signatures are not scale invariant, as they scale according to level. Scaling a path $X \mapsto cX$ scales each coordinate-wise derivative scales by $c$, so level 1 signatures scale likewise. However, level 2 signatures have $(c\dot{x}_i) (c\dot{x}_j)$, so they scale as $c^2$. In general, level $m$ signatures scale as $c^m$.

Scaling therefore does not multiply the kernel by a constant: $K(cX,cY)=\sum_m c^{2m}\sum_{|U|=m}S^U(X) S^U(Y)$ reweights the levels against each other. To compare shapes regardless of their size, we normalize every path to the same total variation (Euclidean length) before computing any kernel.

We normalize by total variation rather than by the $\ell^2$ norm of the increments, $\big(\sum_i\|X(t_{i+1})-X(t_i)\|^2\big)^{1/2}$, because it does not depend on curve sampling. Take a single segment of length 1 and add its midpoint: the total variation is $\tfrac12+\tfrac12 = 1$ both times, while the sum of squared increments drops from $1$ to $\tfrac14+\tfrac14=\tfrac12$. Normalizing by the latter would assign different scales to the same curve depending on how many points it was sampled at, which breaks reparametrization invariance. We also use the Euclidean rather than the $\ell^1$ total variation because the $\ell^1$ length changes when a path is rotated.

Choosing the length is less simple than for SRVF, where every curve is scaled to norm $1$. Normalizing to length $1$ can make the higher-level signatures indistinguishable. By the per-level bound above, the contribution of level $m$ is at most $(L^m/m!)^2$, which is largest for $m$ close to $L$, so for small $L$ the higher levels become small very quickly. We also cannot make paths very long, as the kernel grows exponentially in the length and this leads to numerical instability. We report our results for several lengths, chosen in Section~\ref{sec:parameters}.

\subsection{Readouts}\label{sec:readouts}

We also consider alternatives to using just $S_{[0,1]}(X)$ to compare paths. The PDE is posed on the whole square $[0,1]^2$, so solving it gives $K_{(s,t)}(X,Y)$ for every pair of prefix paths. As solving for $K_{(1,1)}(X,Y)$ requires solving for the entire grid, reading more of it costs no extra solves. This leaves two independent choices: which part of the square we read, and what we compare.

\paragraph{Readout regions.} We read either the \emph{corner} $(1,1)$, the \emph{diagonal} $\{(a,a) : a\in[0,1]\}$, or the full \emph{surface} $[0,1]^2$. Write $\Omega$ for the region and $\mu$ for its natural measure (a point mass, $\diff a$ along the diagonal, or $\diff s \diff t$).

\paragraph{Comparison methods.} Given a region, there are two ways to turn the square into a comparison between $X$ and $Y$.

\begin{itemize}
\item \emph{Signature method.} Solve $X$ against $Y$ and integrate the cross square over the region: \[
  k^{\mathrm{sig}}_\Omega(X,Y)=\int_\Omega K_{(s,t)}(X,Y) \diff \mu(s,t).
  \] On the diagonal this is \[
  k^{\mathrm{sig}}_{\mathrm{diag}}(X,Y)=\int_0^1 \langle S_{[0,a]}(X),S_{[0,a]}(Y)\rangle \diff a,
  \] the $\mbL^2([0,1];H)$ inner product of the running signatures $a\mapsto S_{[0,a]}(X)$ and $a\mapsto S_{[0,a]}(Y)$, where $H$ is the Hilbert space in which signatures take values. The two paths are compared through their signatures at every matching time, not only at the end.
\item \emph{Function-embedding method.} Solve each path against itself, giving the real-valued surface $G_X(s,t):=K_{(s,t)}(X,X)$, which depends on $X$ alone. Restricting $G_X$ to the region embeds $X$ in the function space $\mbL^2(\Omega,\mu)$, and we compare paths with the usual function inner product: \[
  k^{\mathrm{fun}}_\Omega(X,Y)=\int_\Omega G_X(s,t) G_Y(s,t) \diff \mu(s,t).
  \]
\end{itemize}

The signature method is linear in one cross square, while the function-embedding method is bilinear in two self-squares. This is easy to blur on the diagonal, since both produce a real-valued function of $a$: $K_{(a,a)}(X,Y)$ compares the pair, while $K_{(a,a)}(X,X)$ is a feature of $X$ alone.

Table~\ref{tab:readouts} lists the six combinations. Not every combination is useful, as each method can degenerate.

\begin{table}[t]
\centering
\renewcommand{\arraystretch}{1.5}
\caption{The kernel $k(X,Y)$ given by each readout region and comparison method. The two degenerate combinations are not used.}
\begin{tabular}{@{}l>{\raggedright\arraybackslash}p{0.43\linewidth}>{\raggedright\arraybackslash}p{0.37\linewidth}@{}}
\toprule
Region & Signature method & Function-embedding method \\
\midrule
Corner & $K(X,Y)$, the signature kernel & $\|S(X)\|^2\|S(Y)\|^2$, only norms (degenerate) \\
Diagonal & $\int_0^1\langle S_{[0,a]}(X),S_{[0,a]}(Y)\rangle \diff a$ & $\int_0^1\|S_{[0,a]}(X)\|^2\|S_{[0,a]}(Y)\|^2 \diff a$ \\
Surface & $\langle\int_0^1 S_{[0,s]}(X) \diff s,\int_0^1 S_{[0,t]}(Y) \diff t\rangle$, the mean signature (degenerate) & $\int_0^1\int_0^1 G_X G_Y \diff s \diff t$ \\
\bottomrule
\end{tabular}

\label{tab:readouts}
\end{table}

The signature surface collapses because integrating $s$ and $t$ separately throws away which time of $X$ is matched with which time of $Y$, leaving only the time-averaged signature of each path. A genuinely two-parameter signature feature would be the interval signature $(s,t)\mapsto S_{[s,t]}(X)$, that would require solving a quadratic number of PDEs, one for each starting sample, and thus would be computationally infeasible. Besides the standard corner kernel, we use the \emph{diagonal signature} readout and the \emph{diagonal} and \emph{surface function-embedding} readouts.

Further, all of the chosen readouts are still bounded positive semi-definite (PSD) kernels. On the signature side of the table, an integral of PSD kernels is still PSD, and we can obtain the bound by recognizing that subpaths are shorter than the full path, and applying the kernel bound $|K_{(1,1)}(X,Y)|\leq e^{L_{X}+L_{Y}}$ to obtain an integral of a bounded function over a bounded domain, making the whole operation bounded. On the function-embedding side of the table, we recognize this as a separate feature map, and the kernel as the inner product under this feature map, making it PSD and bounded.

Both of these read values at interior times $s,t$, so unlike the corner they depend on how the paths are parametrized. For example, take the straight segment from $0$ to $e_1$, traversed at constant speed by $Y(t)=t e_1$ and a different parametrization $X(t)=t^2e_1$. Both trace the same curve, so their full signatures agree at $t=1$. At $a=\tfrac12$, however, the diagonal compares the prefix of $X$ ending at $\tfrac14e_1$ with the prefix of $Y$ ending at $\tfrac12e_1$. The diagonal would report a difference between two traversals of the same curve.

\paragraph{Arc-length reparametrization.} We remove this dependence by re-timing every path to constant speed before reading the square: at time $a$, the re-timed path $\bar{X}$ has covered exactly the fraction $a$ of its length. In the example, $X$ has covered the fraction $t^2$ of its length by time $t$, so $\bar{X}(a)=X(\sqrt{a})=a e_1=Y(a)$. In general, the fraction of the length already covered when the path reaches a given point of the curve does not depend on how fast it got there, so $X$ and any reparametrization $X\circ\psi$ have the same re-timed path. Every readout computed on $\bar{X}$ and $\bar{Y}$ is therefore reparametrization invariant, and the diagonal at $a$ compares the prefixes of the two paths that cover the same fraction $a$ of their lengths. This re-timing acts on time, not on space, so it is independent of the length normalization above. From here on we assume all paths have been re-timed, and drop the bar.

\paragraph{Rotation.} Signatures are not rotation invariant. Level-1 signatures are the increments $S^{i}_{[0,1]}(X)=X_i(1)-X_i(0)$, so rotating a path by $O\in SO(3)$ rotates them along with it, and the level-1 term of $K(X,OY)$, $\langle X(1)-X(0), O(Y(1)-Y(0))\rangle$, changes with $O$. Rotating both paths together, however, changes nothing. In the PDE of Theorem~\ref{thm:goursat}, the paths enter only through $\langle\dot X(s),\dot Y(t)\rangle$, and $\langle O\dot X(s),O\dot Y(t)\rangle=\langle\dot X(s),\dot Y(t)\rangle$. Both kernels solve the same PDE with the same boundary values, so $K_{(s,t)}(OX,OY)=K_{(s,t)}(X,Y)$ for every $(s,t)$. In particular, the surface function-embedding readout is already rotation invariant, since $G_{OX}(s,t)=K_{(s,t)}(OX,OX)=G_X(s,t)$. It needs no alignment and remains a PSD kernel.

For the signature-method readouts (corner and diagonal) we handle rotation as in Section~\ref{sec:elastic-shape-analysis}, by aligning the pair before comparing them. We align over $SO(3)$ rather than $O(3)$, since a reflection turns a chromosome into its mirror image, which is a physically different conformation. Let $k$ be either readout. Being PSD, it induces the distance $d_k(X,Y)^2=k(X,X)+k(Y,Y)-2k(X,Y)$, and we set \[
d(X,Y)=d_k(X,O^{*}Y),\qquad O^{*}=\arg\min_{O\in SO(3)}\|X-OY\|_2,
\] where $O^{*}$ is the Procrustes rotation of the points of $Y$ onto those of $X$. SRVF starts from the same alignment and then refines it, alternating rotations of the SRVFs with the optimization over reparametrizations. By the invariance above, $k(O^{*}Y,O^{*}Y)$ equals $k(Y,Y)$, so only the cross term is affected. Because $O^{*}$ is chosen separately for every pair, the matrix of aligned kernel values is in general no longer positive semi-definite, and after alignment we no longer have a kernel. We instead lower it to a distance on the quotient of paths by rotations, which is all the energy test of Section~\ref{sec:statistical-testing} requires.

The surface function-embedding readout has a problem of scale. By the bound of Section~\ref{sec:invariances}, applied with Euclidean norms, the level-$m$ signature of a path $X$ of length $L_X$ has norm at most $L_X^m/m!$, with equality when the path is a straight segment. Since every prefix of an arc-length reparametrized path of length $L$ ending at time $a$ has length $aL$, summing over levels gives
\[
G_X(a,a)=\|S_{[0,a]}(X)\|^2\le\sum_{m\ge0}\frac{(aL)^{2m}}{(m!)^2}=I_0(2aL),
\]
where $I_0$ is the modified Bessel function of the first kind. This bound grows like $e^{2aL}$, and for the lengths we use the self-surface $G_X$ is many orders of magnitude larger near the corner $(1,1)$ than anywhere else. The integral $\int G_X G_Y$ is then dominated by that corner. The surface readout therefore behaves like the degenerate corner entry of the table, $\|S(X)\|^2\|S(Y)\|^2$, which only compares the sizes of the two signatures.

We call the readouts built from $G_X$ itself the raw function embeddings, and remove this growth in two ways. The first divides out the diagonal of the surface itself,
\[
\widehat G_X(s,t)=\frac{G_X(s,t)}{\sqrt{G_X(s,s) G_X(t,t)}},
\]
which is the cosine of the angle between the signatures of the prefixes ending at $s$ and $t$. Its diagonal is identically $1$, so we only use it on the surface. The second divides by the bound above,
\[
\overline G_X(s,t)=\frac{G_X(s,t)}{\sqrt{I_0(2sL) I_0(2tL)}}.
\]
By the Cauchy--Schwarz inequality and the bound, $|\overline G_X|\le1$, and on the diagonal $\overline G_X(a,a)=1$ exactly when the prefix ending at $a$ is straight. The diagonal therefore measures how far each prefix is from a straight segment, so we use $\overline G_X$ on both the surface and the diagonal. In both cases the normalizer depends only on the path itself or only on the position $(s,t)$, so the normalized readouts are still inner products of features of single paths. They remain bounded PSD kernels, and like $G_X$ they are rotation invariant and need no alignment.

\section{Statistical Testing}\label{sec:statistical-testing}

For each region we have $60$ reconstructed conformations in each of three conditions, WT, KO and CKO. The conformations of a condition are all reconstructed from that condition's Hi-C contact matrix. We compare WT with KO and WT with CKO. A comparison takes the samples $X_1,\dots,X_{n_1}$ and $Y_1,\dots,Y_{n_2}$ of two conditions, with $n_1=n_2=60$ and $N=n_1+n_2=120$, and asks whether both come from the same distribution of shapes. The test only sees the matrix of pairwise distances between the $N$ conformations, which can be the SRVF distance of Section~\ref{sec:elastic-shape-analysis} or one of the signature distances of Section~\ref{sec:readouts}.

The null hypothesis is that both samples are drawn from one common distribution, which means the labels would carry no information. If we pool the $N$ conformations and hand out the labels again at random, keeping $n_1$ of one and $n_2$ of the other, the new data set has the same distribution as the observed one. We use the method outlined in \citet{hagwoodTestingEqualityCell2013}, using their energy statistic \[
\mathcal{E}=\frac{n_1n_2}{N}\left(\frac{2}{n_1n_2}\sum_{i=1}^{n_1}\sum_{j=1}^{n_2}d(X_i,Y_j)
-\frac{1}{n_1^2}\sum_{i,i'=1}^{n_1}d(X_i,X_{i'})
-\frac{1}{n_2^2}\sum_{j,j'=1}^{n_2}d(Y_j,Y_{j'})\right).
\] This statistic compares the average distance between the two samples with the average distances within each of them. When both samples come from the same distribution these averages agree and $\mathcal{E}$ is close to zero. When conformations of the same condition are closer to each other than to those of the other condition, $\mathcal{E}$ is large. Large values of $\mathcal{E}$ are therefore evidence against the null.

The distribution of $\mathcal{E}$ under the null depends on the unknown distribution of the conformations, so we use a permutation test \citep{van2000asymptotic} to simulate it. We draw $B$ random relabelings of the pooled conformations, each with $n_1$ conformations in the first sample, and compute the statistic $\mathcal{E}_b$ on each. With this permutation test we extract the $p$-value \[
p=\frac{1+\#\{b : \mathcal{E}_b\ge \mathcal{E}\}}{1+B},
\] where the observed labeling counts as one of the $1+B$ labelings. A relabeling only permutes the rows and columns of the distance matrix, so no distance has to be recomputed. We use $B=10^4$ throughout, so the smallest possible $p$-value is about $10^{-4}$.

A test on a whole region tells us whether the conformations differ somewhere along the chromosome, but not where. To locate the differences we also test each topologically associating domain (TAD) on its own. The Hi-C matrix divides the chromosome into consecutive bins and records how often each pair of bins is found in contact. Every point of a conformation corresponds to one bin. A TAD is a stretch of bins that contact each other much more often than they contact bins outside it. TADs therefore appear as bright blocks along the diagonal of the matrix, separated by places that few contacts cross. We find these places with the insulation score, as in \citet{schwarzer2017two}. At each bin, the insulation score averages the contacts between a window of bins just before it and a window just after it. The score is high inside a TAD and drops at a boundary, so we take the TAD boundaries to be the pronounced local minima of the score.

We call the boundaries on the WT matrix and use the same boundaries for all three conditions, so that a TAD covers the same stretch of the chromosome in every comparison. Each TAD is then cut out of every conformation and resampled to a fixed number of points by arc length. Together with the length normalization of Section~\ref{sec:invariances}, this means the test compares the shapes of the TADs and not how many bins they span. Each TAD is tested in the same way as a whole region.

Testing every TAD gives one $p$-value per TAD, and with many TADs in a region some will be small by chance alone. Within each region, comparison and distance, we therefore control the false discovery rate, the expected fraction of false rejections among the rejected TADs, with the Benjamini--Hochberg procedure \citep{benjamini1995fdr}. The tests of neighboring TADs are not independent, but the procedure remains valid under the positive dependence we expect between them \citep{benjamini2001dependency}.

We also check the pipeline by doing a null calibration. Within a single condition, we split the $60$ conformations at random into two halves and test one half against the other, a situation in which the null is true. Both halves are reconstructed from the same Hi-C matrix and the labels are assigned at random, so every rejection is a false positive. We repeat this over many random splits for every region, TAD, distance and condition. For a calibrated test, the fraction of splits rejected at level $\alpha$ is a binomial proportion with mean $\alpha$, and the $p$-values are uniformly distributed. Since the labels are random, the test is exact here regardless of the distance, so this check verifies the implementation of the test and not the quality of the distances. It says nothing about the comparisons between conditions, whose conformations come from different Hi-C matrices.

\section{Experimental Results}
\subsection{Sensitivity to noise}\label{sec:noise}

\begin{figure}[t]
\centering
\includegraphics[width=\textwidth]{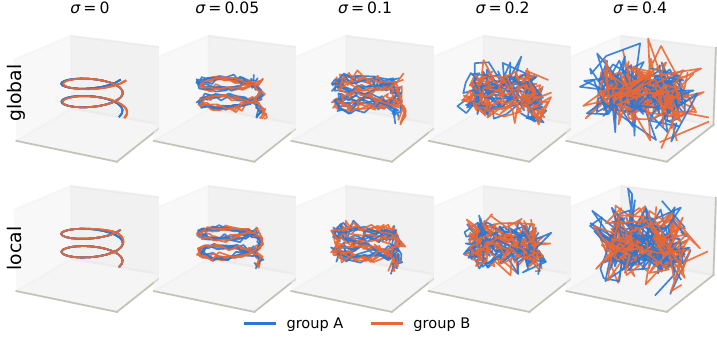}
\caption{Four curves of group A (blue) and of group B (orange) from one data set with $\delta=0.2$, for the global (top) and local (bottom) deformation and increasing noise $\sigma$.}
\label{fig:noise-gallery}
\end{figure}

To see how the methods respond to noise, we compare them on synthetic curves for which both the difference between the groups and the amount of noise are known. The base curve is a helix of radius $1$ and height $1$ with two turns, sampled at $50$ equally spaced values of $t\in[0,1]$. Group A consists of copies of the base helix, and group B of helices deformed by an amount $\delta$, in one of two ways. The global deformation bends the whole curve by adding $4\delta(t-1/2)^2$ to its first coordinate, so the two ends move by $\delta$ and the middle of the curve stays in place. The local deformation adds a narrow bump $\delta\exp(-((t-0.6)/0.05)^2)$ to the second coordinate, which changes only a few points around $t=0.6$. The size of the deformation varies by $10\%$ around $\delta$ from curve to curve, so that group B is a distribution of shapes and not a single shape. Finally, we add independent Gaussian noise with standard deviation $\sigma$ to every coordinate of every point of every curve in both groups. Figure~\ref{fig:noise-gallery} shows examples of both deformations for $\delta=0.2$ and increasing $\sigma$.

Each group has $60$ curves, the sample size of the comparisons between conditions, and the $50$ points per curve match the length to which we resample each TAD in the per-TAD tests. We vary the deformation over $\delta\in\{0,0.1,0.2,0.5\}$ and the noise over $\sigma\in\{0,0.02,0.05,0.1,0.2,0.4\}$. For each kind of deformation and each pair $(\delta,\sigma)$ we draw $100$ independent data sets, run the energy test of Section~\ref{sec:statistical-testing} on each of them with $B=10^4$ permutations, and record the fraction of data sets in which the test rejects at level $0.05$. When $\delta=0$ both groups have the same distribution, and this fraction estimates the probability of a false rejection, which should be close to $0.05$. When $\delta>0$ it estimates the power of the test, the probability of detecting the deformation. With $100$ data sets the standard error of each fraction is at most $0.05$.

We compare SRVF, aligned over $SO(3)$, with four of the signature readouts of Section~\ref{sec:readouts}: the corner, the diagonal, the self-cosine surface and the straight-line normalized surface, each with the distance $d_k$ defined there. We also run SRVF on smoothed curves, where each coordinate of each curve is convolved with a Gaussian of standard deviation two sample spacings, and we return to this at the end of the section. The signature readouts are computed at total variation $L\in\{4,8,16,32\}$. For the chromosome data we only use $L\geq 8$, because at smaller $L$ the kernel values of different conformations agree in nearly all of their digits and the distances are lost to cancellation (Section~\ref{sec:parameters}). On the synthetic curves the distances at $L=4$ stay well away from zero and the test keeps its level, as we show next, so we include $L=4$ here.

When $\delta=0$ and $\sigma>0$ the two groups have the same distribution, and every method rejects at close to the nominal rate. Over all methods, both deformations and the five noise levels, the fraction of rejected data sets is $0.049$ on average and at most $0.09$, which is within the variation expected from $100$ data sets. We leave out the setting $\delta=\sigma=0$, in which every curve of both groups is the same helix. All distances are then zero up to rounding error, so the rejection rate in this setting says nothing about the methods.

\begin{figure}[t]
\centering
\includegraphics[width=\textwidth]{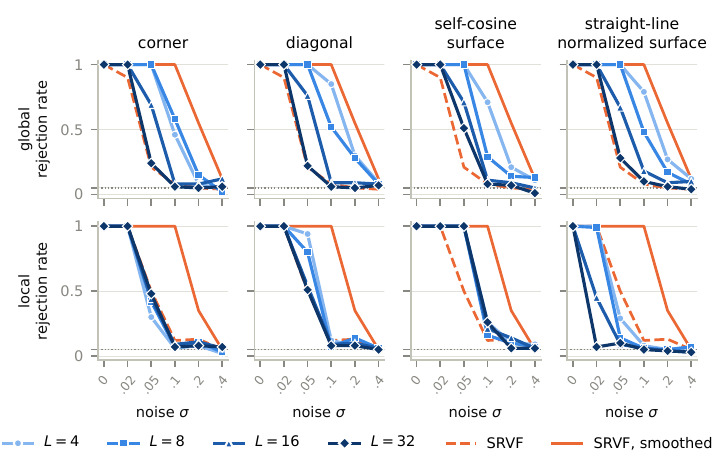}
\caption{Fraction of the $100$ data sets in which the test rejects at level $0.05$, against the noise $\sigma$, for $\delta=0.1$, with the global deformation in the top row and the local deformation in the bottom row. Blue lines are the signature readouts at $L=4,8,16,32$, darker for larger $L$, and orange lines are SRVF on the original (dashed) and smoothed (solid) curves. The dotted line marks $0.05$.}
\label{fig:noise-power-01}
\end{figure}

\begin{figure}[t]
\centering
\includegraphics[width=\textwidth]{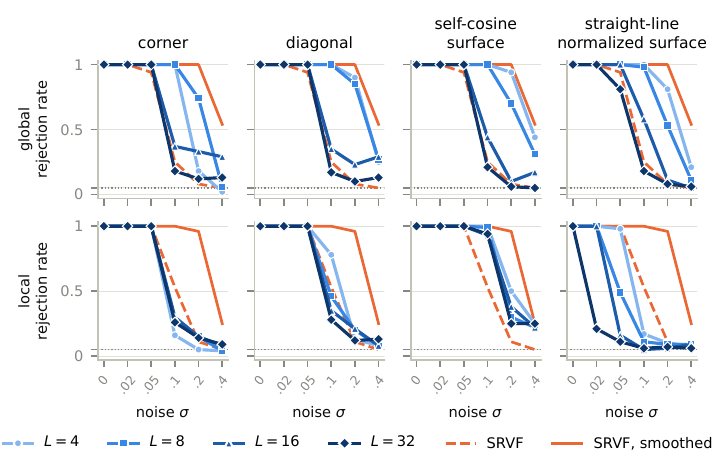}
\caption{As Figure~\ref{fig:noise-power-01}, for $\delta=0.2$.}
\label{fig:noise-power-02}
\end{figure}

Figures~\ref{fig:noise-power-01} and~\ref{fig:noise-power-02} show how the fraction of rejected data sets changes as the noise grows, for $\delta=0.1$ and $\delta=0.2$. Without noise every method detects both deformations in every data set, and at $\sigma=0.02$ nearly all of them still do. For $\delta=0.5$ every method except the straight-line normalized surface at $L\geq 8$ detects the deformation in at least $90\%$ of the data sets up to $\sigma=0.1$, so we do not show this case. As the noise grows, the methods lose power at different rates. For the global deformation with $\delta=0.2$ and $\sigma=0.1$, SRVF rejects in $25\%$ of the data sets, while all four readouts reject in at least $98\%$ at $L=4$ and $L=8$.

For the global deformation the power of the signature readouts depends on $L$, and the readouts keep their power at higher noise levels for smaller $L$. In the setting above the readouts reject in $35\%$ to $58\%$ of the data sets at $L=16$, and in $17\%$ to $21\%$ at $L=32$, which is close to SRVF. For the local deformation the differences between the readouts are larger than those between values of $L$. The corner and diagonal readouts do not keep their power longer than SRVF, except for the diagonal readout at small $L$. For $\delta=0.2$ and $\sigma=0.1$, SRVF rejects in $53\%$ of the data sets, the corner readout in $16\%$ to $30\%$, and the diagonal readout in $28\%$ at $L=32$ up to $78\%$ at $L=4$. In the same setting the self-cosine surface rejects in at least $93\%$ of the data sets at every $L$, the highest power of all methods applied to the original curves. The straight-line normalized surface loses the local deformation at the lowest noise levels. At $L\geq 8$ its power falls below $50\%$ already at $\sigma=0.05$, and at $L=32$ already at $\sigma=0.02$.

Smoothing the curves before computing SRVF recovers most of the power it loses to noise. With the smoothing described above, SRVF detects the global deformation with $\delta=0.2$ in every data set up to $\sigma=0.2$, where it rejects in $8\%$ of the data sets without smoothing, and the local deformation with $\delta=0.2$ in $96\%$ of the data sets at $\sigma=0.2$, compared with $11\%$ without smoothing. Its power is then at least as high as that of the best signature readout in every setting except two at $\sigma=0.4$, in which no method rejects in more than $13\%$ of the data sets. SRVF is computed from the derivative of the curve, which amplifies the noise added to each point, and smoothing removes most of this noise before the derivative is taken. We did not smooth the curves before computing the signature readouts.

\subsection{Data}\label{sec:data}
For our data we have chromosome conformations estimated from High-throughput Chromosome (Hi-C) Conformation Capture bulk data \cite{lieberman2009comprehensive}. Hi-C bulk data comes in the form a $n\times n$ matrix where the entries capture interactions between regions. That is, the larger the $ij$ entry of the matrix, the more interactions are observed between the $i$th and $j$th regions and hence the inference is that these regions are spatially nearby. The size of the regions is a function of the resolution, or bin size, of the capturing method. There is, however, a trade-off as the capturing technique is noisy, so the smaller the bin size the more noise\cite{oluwadare2019overview}.

There is, however, no consensus on the correct method for estimating a conformation from Hi-C data, as this is an inverse problem with many possible local solutions.
We utilize SIMBA3D \cite{rosenthal2019bayesian} as we are comparing our methods to \cite{soto2021statistical}. The conformations are smooth, continuous three-dimensional curves. We aim to compare these curves in their entirety and also at localized areas, so we require a method to segment these curves. Rather than do a uniform segmentation, we utilize TADs \citep{dixon2012topological}. TADs are regions of the chromosome which are highly associated with each other, for instance loops.

The matrices come from three conditions, wild type (WT) and two knock-outs, KO and CKO, in which the genes RIF1 and RAD21 respectively have been knocked out. Wild type data is control while knock out type data have had a gene forced to be inactive or unable to be expressed. As structure and function are intimately related, the hypothesis is that the conformations estimated from the KO and CKO data have shape differences at a local or global level, or both. 

\subsection{Parameters}\label{sec:parameters}

For the per-TAD tests we aim to reproduce the segmentation of \citet{soto2021statistical}, who report $46$ TADs for one of their regions. We compute the insulation score with a window of $10$ bins on each side, which gives a comparable number of TADs per region. We call a boundary at every local minimum of the $\log_2$ score with prominence at least $0.1$, and require at least $15$ bins between consecutive boundaries. The minimum size is important as every TAD is resampled to $50$ points: a very short TAD is upsampled several times over, and in our runs the shortest TADs had inflated values of $\mathcal{E}$. This gives $339$ TADs, between $40$ and $44$ per region. They span between $10$ and $52$ bins, with a median of $22$; the shortest ones are the first and last TAD of a region, which are not subject to the minimum spacing.

We compute the signature readouts at total variation $L\in\{8,16,32\}$. Both ends of this range are limited by floating-point arithmetic. When $L$ is too small, the higher levels are suppressed and the kernel values of different paths agree in nearly all of their digits, so the distances, which subtract them, are lost to catastrophic cancellation. When $L$ is too large, the kernel, which can grow like $I_0(2L)$, overflows. We found the range above to avoid both. For SRVF we align over $SO(3)$. Every permutation test uses $B=10^4$ relabelings, and we control the false discovery rate at level $0.05$.

\subsection{Calibration}

Each calibration cell uses $500$ random splits of the $60$ conformations of a condition into two halves of $30$. We check three things. First, the fraction of splits with $p<0.05$, averaged over cells, should be close to $0.05$. Second, the rejection rate of a single cell is itself noisy, and for a calibrated test it falls in $[0.032,0.070]$ in $95\%$ of cells, so about $5\%$ of cells should fall outside this band. Third, the $500$ $p$-values of a cell should be uniform on $[0,1]$, which we check with a Kolmogorov--Smirnov test that should also reject in about $5\%$ of cells.

All three checks agree with a calibrated test. On whole regions, the average rejection rate is $0.050$, $2.9\%$ of cells fall outside the band and $4.7\%$ fail the uniformity test. Per TAD, the average rejection rate of every distance is between $0.049$ and $0.051$, between $2.6\%$ and $5.9\%$ of its cells fall outside the band, and between $3.3\%$ and $6.8\%$ fail the uniformity test.

\subsection{Whole regions}\label{sec:whole}

\begin{table}[t]
\centering
\caption{Number of regions, out of $8$, in which the test rejects at level $0.05$.}
\begin{tabular}{lcccccc}
\toprule
 & \multicolumn{3}{c}{WT vs KO} & \multicolumn{3}{c}{WT vs CKO} \\
\cmidrule(lr){2-4}\cmidrule(lr){5-7}
 & $L=8$ & $16$ & $32$ & $8$ & $16$ & $32$ \\
\midrule
SRVF & \multicolumn{3}{c}{8} & \multicolumn{3}{c}{8} \\
Corner & 3 & 3 & 3 & 7 & 8 & 8 \\
Diagonal & 6 & 6 & 6 & 8 & 8 & 8 \\
Raw surface function embedding & 8 & 8 & 8 & 7 & 8 & 8 \\
Raw diagonal function embedding & 8 & 8 & 8 & 8 & 8 & 8 \\
Self-cosine surface & 8 & 8 & 8 & 8 & 8 & 8 \\
Straight-line normalized surface & 7 & 8 & 8 & 8 & 8 & 8 \\
Straight-line normalized diagonal & 8 & 7 & 8 & 8 & 8 & 8 \\
\bottomrule
\end{tabular}

\label{tab:whole}
\end{table}

Table~\ref{tab:whole} counts, for each method, the regions in which the test rejects at level $0.05$. For WT against CKO every method rejects in all eight regions at every length, except the corner readout and the raw surface function embedding at $L=8$, which reject in seven. WT against KO separates the methods. SRVF rejects in all eight regions, all at the smallest attainable $p$-value. The two signature methods are the least sensitive: the corner readout rejects in three regions and the diagonal readout in six. The function embeddings reject in seven or eight regions. The self-cosine surface rejects in all eight regions at every length and, like SRVF, always at the smallest attainable $p$-value. We return to the raw function embeddings in Section~\ref{sec:collapse}, as their rejections turn out to measure mostly the size of the signature.

\subsection{TADs}\label{sec:tad}

\begin{table}[t]
\centering
\caption{Per-TAD results. For each comparison, the columns give the number of TADs out of $339$ rejected after Benjamini--Hochberg, the number of TADs rejected by SRVF that are also rejected, and the Spearman correlation $\rho$ of $\mathcal{E}$ with SRVF over TADs.}
\begin{tabular}{lcccccc}
\toprule
 & \multicolumn{3}{c}{WT vs KO} & \multicolumn{3}{c}{WT vs CKO} \\
\cmidrule(lr){2-4}\cmidrule(lr){5-7}
$L$ & rejected & of SRVF's $222$ & $\rho$ & rejected & of SRVF's $332$ & $\rho$ \\
\midrule
SRVF & 222 & 222 & 1.00 & 332 & 332 & 1.00 \\
\midrule
\multicolumn{7}{l}{\emph{Corner}} \\
8 & 286 & 207 & 0.44 & 339 & 332 & 0.39 \\
16 & 318 & 217 & 0.39 & 339 & 332 & 0.33 \\
32 & 325 & 218 & 0.38 & 339 & 332 & 0.32 \\
\midrule
\multicolumn{7}{l}{\emph{Diagonal}} \\
8 & 272 & 200 & 0.48 & 339 & 332 & 0.42 \\
16 & 317 & 216 & 0.39 & 339 & 332 & 0.34 \\
32 & 325 & 218 & 0.38 & 339 & 332 & 0.33 \\
\midrule
\multicolumn{7}{l}{\emph{Raw surface function embedding}} \\
8 & 267 & 190 & 0.39 & 338 & 331 & 0.35 \\
16 & 220 & 162 & 0.35 & 336 & 329 & 0.31 \\
32 & 229 & 166 & 0.35 & 335 & 328 & 0.31 \\
\midrule
\multicolumn{7}{l}{\emph{Raw diagonal function embedding}} \\
8 & 256 & 185 & 0.38 & 337 & 330 & 0.35 \\
16 & 221 & 161 & 0.35 & 336 & 329 & 0.31 \\
32 & 230 & 167 & 0.35 & 334 & 327 & 0.31 \\
\midrule
\multicolumn{7}{l}{\emph{Self-cosine surface}} \\
8 & 330 & 219 & 0.42 & 339 & 332 & 0.42 \\
16 & 335 & 220 & 0.50 & 339 & 332 & 0.52 \\
32 & 337 & 221 & 0.55 & 339 & 332 & 0.54 \\
\midrule
\multicolumn{7}{l}{\emph{Straight-line normalized surface}} \\
8 & 298 & 208 & 0.49 & 338 & 331 & 0.44 \\
16 & 289 & 204 & 0.47 & 339 & 332 & 0.42 \\
32 & 291 & 203 & 0.47 & 339 & 332 & 0.43 \\
\midrule
\multicolumn{7}{l}{\emph{Straight-line normalized diagonal}} \\
8 & 289 & 204 & 0.46 & 338 & 331 & 0.42 \\
16 & 290 & 202 & 0.45 & 339 & 332 & 0.41 \\
32 & 287 & 200 & 0.44 & 339 & 332 & 0.41 \\
\bottomrule
\end{tabular}

\label{tab:tad}
\end{table}

Table~\ref{tab:tad} gives the per-TAD results. For WT against CKO the comparison is saturated: SRVF rejects $332$ of the $339$ TADs, the signature methods and the normalized function embeddings reject $338$ or $339$, including all or all but one of the TADs that SRVF rejects. The raw function embeddings reject between $334$ and $338$.

For WT against KO, SRVF rejects $222$ TADs, and every signature-based method rejects more. The self-cosine surface rejects between $330$ and $337$. The corner and diagonal readouts reject between $272$ and $325$, more at larger $L$, and the straight-line normalized surface and diagonal between $287$ and $298$. These larger sets largely contain the TADs that SRVF rejects. The self-cosine surface rejects $219$ to $221$ of SRVF's $222$, the corner and diagonal readouts $200$ to $218$, and the straight-line normalized readouts $200$ to $208$. The additional TADs number between $72$ and $116$, depending on the method. The methods therefore mostly agree on which TADs change, and the signature-based methods add TADs that SRVF does not reject. The ranking of the TADs by $\mathcal{E}$ agrees less: its Spearman correlation with the ranking by SRVF is between $0.38$ and $0.48$ for the corner and diagonal readouts and between $0.42$ and $0.55$ for the normalized function embeddings. The raw function embeddings agree least with SRVF, covering as few as $161$ of its TADs with correlations between $0.35$ and $0.39$, which we explain in Section~\ref{sec:collapse}.

\subsection{Function embeddings}\label{sec:collapse}

The growth of the self-surface described in Section~\ref{sec:readouts} is large enough in our data to make the raw function embeddings degenerate. On the first six TADs of region 1, the corner square covering the last $10\%$ of the surface in each direction holds $41\%$, $93\%$ and $99.9\%$ of the squared norm of the surface at $L=8$, $16$ and $32$, and the surface spans $3$, $10$ and $24$ orders of magnitude. The integral defining the readout is then essentially a product of the two corner values. Across all $339$ TADs, the cosine $\hat k(X,Y)=k(X,Y)/\sqrt{k(X,X) k(Y,Y)}$ between two conformations has median $0.977$ to $0.9998$ for the raw function embeddings, so their features are nearly parallel.

To remove this dominance we normalize the self-surfaces before comparing them, as described in Section~\ref{sec:readouts}: either by the diagonal of the surface itself, which gives the self-cosine surface $\widehat G_X$, or by the bound $I_0$ for a straight path, which gives the straight-line normalized surface and diagonal $\overline G_X$. Every value of a normalized surface is at most $1$ in absolute value, so no part of the surface dominates the integral. For WT against KO, the number of TADs rejected by a raw function embedding changes by up to $47$ across $L\in\{8,16,32\}$, while for each normalized readout it changes by at most $9$. The normalized readouts also agree better with SRVF, covering $200$ to $221$ of its $222$ TADs instead of as few as $161$, with Spearman correlations of $0.42$ to $0.55$ instead of $0.35$ to $0.39$.

\section{Discussion}\label{sec:discussion}

Our tests are calibrated, so the TADs they reject are not false positives in the usual sense. What a rejection means, however, depends on the null hypothesis we are testing. All conformations of a condition are reconstructed from the same Hi-C matrix, so when we compare two conditions we are asking whether two sets of reconstructions, each coming from its own matrix, have the same distribution of shapes. A rejection tells us that a method can distinguish these two sets of reconstructions, and the number of TADs a method rejects measures how sensitive the method is, not how many TADs change biologically. In our results the signature-based methods are more sensitive than SRVF, since they reject nearly all of the TADs that SRVF rejects as well as between $72$ and $116$ others, but this does not make their additional rejections biologically significant. With a single Hi-C matrix per condition we have no way of deciding whether these additional TADs reflect a biological change, as this would require several independent matrices per condition.

The methods agree much more on which TADs change than on how strongly they change. While the set of TADs rejected by a signature-based method contains almost all of the TADs rejected by SRVF, the values of the statistic $\mathcal{E}$ are only moderately related. Across TADs, the Spearman correlation between the $\mathcal{E}$ of a signature-based method and the $\mathcal{E}$ of SRVF lies between $0.38$ and $0.48$ for the corner and diagonal readouts, and between $0.42$ and $0.55$ for the normalized function embeddings. The methods therefore detect largely the same changes but weigh their size differently, which is what one would expect from two distances that measure different properties of a curve, its elastic shape on one side and its signature on the other. In particular, a ranking of TADs by $\mathcal{E}$, such as a list of the most strongly changed TADs, depends on the method used to produce it.

On whole regions the results separate the methods less clearly. For WT against KO, SRVF and the self-cosine surface reject in every region with the smallest $p$-value that $10^4$ permutations allow, so at this level the two cannot be told apart. The corner and diagonal readouts, which align each pair of whole conformations with a single rotation, reject in fewer regions.

For WT against CKO every method rejects in nearly every TAD, which leaves little room to compare them. One possible contributor to this saturation is the segmentation. We call the TAD boundaries on the WT matrix and cut the conformations of all three conditions at the same positions, so that each TAD covers the same stretch of the chromosome in every comparison. To see whether these positions are also boundaries for the other conditions, we called boundaries on the KO and CKO matrices with the same procedure, and counted how many WT boundaries have a boundary of the other condition within two bins. For KO this holds for $88\%$ of the WT boundaries, and for CKO for $53\%$, while boundaries placed at random would match about $19\%$. The CKO matrix therefore has domains of its own, but about half of the WT boundaries at which we cut the CKO conformations are not boundaries in the CKO matrix. Many of the stretches we compare against CKO may then cross or split a CKO domain, which could make the two conditions look different in nearly every TAD, although we have not tested whether this explains the saturation.

Our analysis has some limitations, the first of which concerns the function embeddings. Their raw form turned out to compare only the size of each conformation's signature at the lengths we use, which we noticed by examining the surfaces directly. The normalized versions avoid this, but the choices of normalization are arbitrary and chosen by us for practical reasons.

The per-TAD results also depend on the segmentation. We choose the parameters to give a segmentation comparable to that of \citet{soto2021statistical}, and they affect both the stretches of the chromosome that are compared and the number of tests in each region. Using the WT boundaries for every condition also makes the segmentation fit WT and KO better than CKO, as discussed above.

Finally, on the short paths of the per-TAD analysis at $L=32$, the numerical solution of the PDE is not fully converged, and compared with a more accurate solution the kernel values differ by up to $27\%$. The distances we test with are affected much less, since the normalized distances change by at most $0.01$ and the ordering of all pairwise distances stays nearly the same, with a rank correlation of at least $0.998$.

\bibliographystyle{plainnat}
\bibliography{ref}

\end{document}